# ELECTRON CYCLOTRON RESONANCE DISCHARGE AS A SOURCE FOR HYDROGEN AND DEUTERIUM IONS PRODUCTION

Angel.J.Chacon Velasco[1] and V.D.Dougar-Jabon [2]

[1]*Universidad de Pamplona, Pamplona, Colombia*
angelch@unipamplona.edu.co
[2]*Universidad Industrial de Santander, A.A.678 Bucaramanga, Colombia*
vdougar@uis.edu.co

In this report, we describe characteristics of a ring-structure hydrogen plasma heated in electron cyclotron resonance conditions and confined in a mirror magnetic trap and discuss the relative efficiency of secondary electrons and thermoelectrons in negative hydrogen and deuterium ion production. The obtained data and calculations of the balance equations for possible reactions demonstrate that the negative ion production is realized in two stages. First, the hydrogen and deuterium molecules are excited in collisions with the plasma electrons to high-laying Rydberg or vibrational levels in the plasma volume. The second stage leads to the negative ion production through the process of dissociative attachment of low energy electrons. The low energy electrons are originated due to a bombardment of the plasma electrode by ions of one of the driven rings and thermoemission from heated tungsten filaments. Experiments seem to indicate that the negative ion generation occurs predominantly in the limited volume filled with thermoelectrons. Estimation of the negative ion generation rate shows that the main channel of $H^-$ and $D^-$ ion production involves the process of high Rydberg state excitation.

## I. Introduction

Interest to high energy beams of negative hydrogen ions is provoked predominantly by international and national programs of controlled thermonuclear fusion. Plasma sources are acknowledged to be among the most efficient sources for $H^-$ and $D^-$ ion production. It has been found out that the attachment of electrons by excited molecules, $H_2^*$ and $D_2^*$, followed by dissociation of negatively charged molecules, is the main reaction which leads to the formation of respective negative atomic ions. The molecules can be excited to high vibration states when they collide with electrons which have the temperature of (30-50) eV, thus involving Franc-Condon transitions [1,2]. Another type of excited molecules which can participate in negative atomic ion generation are molecules in super excited Rydberg states [3,4]. The transition of molecules to high Rydberg states also occurs at their collisions with electrons. The transition rate reaches a maximum value at the electron temperature of (80-150) eV [5]. Cold electrons contribute efficiently to a dissociative attachment process when their energy does not exceed (3-4) eV. Therefore, the feasibility of plasma negative hydrogen ion production requires two electron components with very different temperatures.

It has been recently pointed out [5,6] that both channels for negative hydrogen ions production are observed in an electron cyclotron resonance (ECR) plasma source, with the reaction channel involving a high Rydberg excitation which is faster than the one involving the vibration excited molecules.

In this report, we present some results obtained on a modified ECR source with driven rings. The modification is aimed at promoting a better rate for hydrogen negative ion generation and consists in designing a system for additional low energy electron supply.

**II. Experimental set up**

Fig.1 shows the schematic diagram of the negative ion source and lines of the magnetic field, discussed in detail elsewhere [5,6]. A discharge is ignited by 2.45 GHz microwaves in an aluminium chamber of 13 cm in diameter and 8 cm long. The chamber is spaced between the magnetic discs which create a mirror type field. The driven rings are formed when the resonance surfaces correspond to the cyclotron electron resonance on the fundamental frequency or its harmonics, $B(r,z) = mc\omega/ek$, take the configuration of a hyperboloid of one sheet [6]. Here $m$ and $e$ are the electron mass and charge respectively, $c$ is the light speed, $\omega$ is the angle frequency of microwaves and $k$ meets the fundamental resonance ($k=1$) and its harmonics ($k=2,3,4$). Fig.1 also shows a cross section of the magnetic surfaces (in the form of lines) and driven plasma rings (in the form of oval circles) by a median *(r,z)*- plane which comprises the chamber axis. The system designated for low energy electrons production consists of 8 tungsten filaments of 0.3 mm in diameter and 3 cm long. They are symmetrically disposed perpendicularly to the magnetic field lines (normally to the plane of Fig.1) with four filaments on the two sides of the extraction orifice. A gap between the filaments and the electrode surface is of 0.6 cm. The current heated filaments emit low energy thermoelectrons in addition to plasma electrons and electrons of secondary emission [5,6]. These electrons diffusing along the magnetic field lines create a layer above the extraction electrode.

**III. Results an discussions**

In the unmodified source [5, 6], the efficient generation of both $H^-$ and $D^-$ ions occurs only when one of the rotating plasma rings touches the plasma electrode. The ions involved in the rotation movement bombard the electrode surface at the angles of 60º-80º with respect to a normal. The bombardment results in a local overheating of the electrode surface which is manifested in coloured strips. On the other hand, the bombardment by high energy ions of a ring causes electron emission from the body of the plasma electrode. The magnetic field prevents these electrons from being diffused into the plasma volume that results in forming a superficial layer of the secondary electrons, so that the two layers of low energy electrons are generated in the immediate vicinity of the extraction orifice. It is in these layers where the excited molecules created in the discharge volume in one way or another can be converted into negatively charged molecules. The negative molecular ions formed in this way are not stable and dissociate easily into excited atoms and negative atomic ions.

Two definite effects of thermoemission of low energy electrons are also observed. First, the gas pressure range which provides a stable discharge regime is found well extended. One can see it in Fig.2 which shows a plasma density dependence on the gas pressure under a fixed microwave power of 175 W for the cases of unheated (curve 1) and the heated filaments (curve 2). These curves make it obvious that the thermoelectrons shift the lower boundary of the stable discharge from $1.7 \times 10^{-4}$ Torr to $0.6 \times 10^{-4}$ Torr and the upper boundary is shifted from $3.3 \times 10^{-4}$ Torr to $5.1 \times 10^{-4}$ Torr. An extension of the working pressure range represents a significant improvement in terms of the source operation. Besides, the termoelectrons provoke a more than twofold increase in the extracted negative ion current (see Fig.3). The peaks on the curves shown in Fig.3 correspond to the cases when the driven rings $k=3$ and $k=4$ touch the plasma electrode surface. On the other hand, Fig.3 shows that the current peaks which are very well marked without thermoelectrons (Curve 1) are found smoothed out with the thermoelectrons (Curve 2), which means that the importance of the driven rings in the negative ion production becomes less significant. Thus, it follows that the contribution of the secondary emitted electrons to the negative ion formation is smaller than the contribution of the thermoelectrons which fill the space between the two sets of filaments. This does not mean that the electron cyclotron discharge can be substituted by another type of discharge. The valuable advantage of the electron cyclotron resonance discharges consists in easy heating of the electron component to the optimal temperature value for excitation of the hydrogen molecules. For the $H^-$ ions produced due to thermoelectrons, a calculated current extracted through a 3-mm orifice under 8 kV voltage is 5.5 mA, while the contribution due to the secondary electrons is estimated as 0.8 mA. It must be noted that the yield of ions $D^-$ is under the 20% of the $H^-$ yield.

Such a suppression of the isotope effect suggests that the survival probabilities of the negative Rydberg molecular ions $H_2^{HR-}$ and $D_2^{HR-}$ do not differ significantly in the characteristic time which required for the molecular nuclei to be moved apart at a distance wherein the excited molecules dissociate with formation or $H^-$ $D^-$ ions and the respective excited atoms. The correlations of the calculated and experimental data are in favour of a higher efficiency of the processes which involve an excitation of superexcited Rydberg states. At the same time, the energetic threshold for a dissociative attachment of the Rydberg excited molecules seems to be lower than the competitive one.

**IV. Conclusion**

The undertaken study has proven that an addition of low energy thermoelectrons to ECR source plasma guarantees an enhancement in the generation of the negative $H^-$ and $D^-$ ions. The secondary electrons due to their relatively high energy are found to be less effective in the processes of negative ion production. This can pave the way for designing new types of ECR sources in which creating conditions for driven plasma ring formation can be disregarded.

The relation between the extracted $H^-$ and $D^-$ currents decreases with addition of the thermoelectrons into the plasma volume. It shows that the main channel for negative ions passes through the high Rydberg states excitation because the isotope effect appears due to the difference in the velocities of nuclear vibration movement in light and heavy hydrogen isotopes.

**Acknowledgements**

The authors wish to thank Dr. O. Bogomolova for valuable assistance. This work is supported by Colciencias under code 1102-05-13566.

6. Dougar-Jabon, V. D., Chacon Velasco, A. J. and Vivas, F. A., Rev. Sci. Instruí. **69**, 950 )1998).

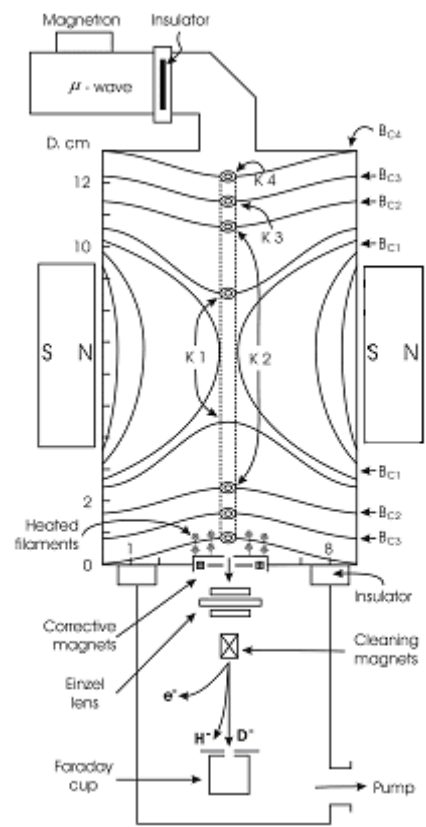

Fig.1. Schematic diagram of the ECR-type negative ion source.

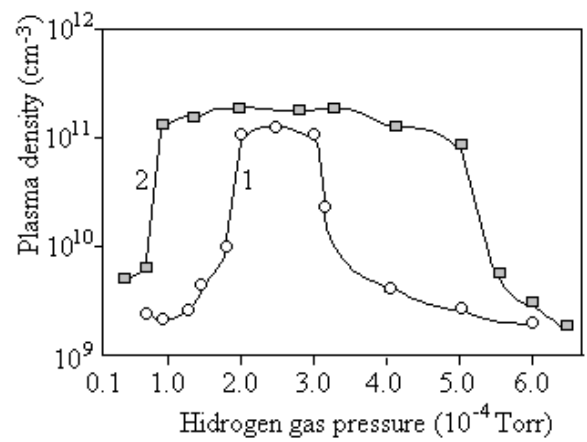

Fig.2. Plasma density dependence of the hydrogen pressure
at the 2.45 GHz microwave power of 200 W. The curves
1 and 2 correspond to unheated and heated tungsten filaments.

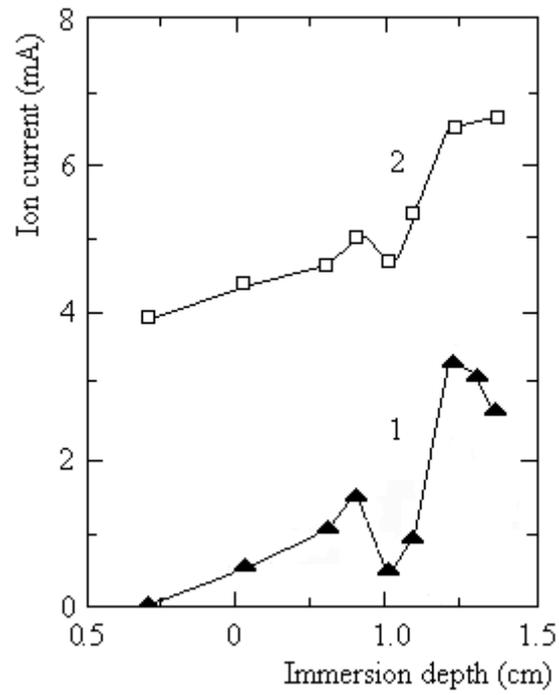

Fig.3. The $H^-$ current as a function of the immersion depth of the plasma electrode into the discharge chamber: a microwave power of 175 W and a pressure of $2 \cdot 10^{-4}$ Torr. The curves 1 and 2 correspond to unheated and heated tungsten filaments.